# Seizure-NGCLNet: Representation Learning of SEEG Spatial Pathological Patterns for Epileptic Seizure Detection via Node-Graph Dual Contrastive Learning

1st Yiping Wang
*School of Biomedical Engineering and Technology,Tianjin Medical University*
Tianjin,China
ypwang@tmu.edu.cn

2nd Peiren Wang
*School of Biomedical Engineering and Technology,Tianjin Medical University*
Tianjin,China
wangpr@tmu.edu.cn

3rd Zhenye Li
*School of Biomedical Engineering and Technology,Tianjin Medical University*
Tianjin,China
zhenyeli@tmu.edu.cn

4th Fang Liu
*Chaoyang Central Hospital, China Medical University*
Chaoyang,China
and_reas6@163.com

5th Jinguo Huang*
*School of Intelligent Engineering and Automation, Beijing University of Posts and Telecommunications*
Beijing,China
hjg@bupt.edu.cn

*Abstract*—Complex spatial connectivity patterns, such as interictal suppression and ictal propagation, complicate accurate drug-resistant epilepsy (DRE) seizure detection using stereotactic electroencephalography (SEEG) and traditional machine learning methods. Two critical challenges remain: (1) a low signal-to-noise ratio in functional connectivity estimates, making it difficult to learn seizure-related interactions; and (2) expert labels for spatial pathological connectivity patterns are difficult to obtain, meanwhile lacking the patterns' representation to improve seizure detection. To address these issues, we propose a novel node-graph dual contrastive learning framework, Seizure-NGCLNet, to learn SEEG interictal suppression and ictal propagation patterns for detecting DRE seizures with high precision. First, an adaptive graph augmentation strategy guided by centrality metrics is developed to generate seizure-related brain networks. Second, a dual-contrastive learning approach is integrated, combining global graph-level contrast with local node-graph contrast, to encode both spatial structural and semantic epileptogenic features. Third, the pretrained embeddings are fine-tuned via a top-k localized graph attention network to perform the final classification. Extensive experiments on a large-scale public SEEG dataset from 33 DRE patients demonstrate that Seizure-NGCLNet achieves state-of-the-art performance, with an average accuracy of 95.93%, sensitivity of 96.25%, and specificity of 94.12%. Visualizations confirm that the learned embeddings clearly separate ictal from interictal states, reflecting suppression and propagation patterns that correspond to the clinical mechanisms. These results highlight Seizure-NGCLNet's ability to learn interpretable spatial pathological patterns, enhancing both seizure detection and seizure onset zone localization.

## I. INTRODUCTION

Approximately 30-40% of epilepsy patients suffer from drug-resistant epilepsy (DRE) [1], for whom surgical intervention remains the primary option to achieve seizure freedom [2, 3]. Precision surgery depends crucially on the accurate identification of the seizure onset zone (SOZ) and the associated epileptic network. When non-invasive methods such as MRI and scalp EEG fail to localize the SOZ[4], clinicians typically turn to stereo-electroencephalography (SEEG), implanting depth electrodes directly into the brain [5, 6], as shown in Fig. 1. Recent studies further suggest that epilepsy is a network-level disorder characterized by complex interactions among brain regions rather than isolated lesions[7, 8]. During the interictal period, the SOZ exhibits increased inward functional connectivity, known as interictal suppression (IS)[9], while this suppressive network may break down in the ictal period, triggering widespread pathological propagation[10]. Therefore, decoding invasive SEEG, with high spatial and temporal resolution, to quantitatively represent interictal suppression and widespread ictal propagation remains a critical scientific challenge[6]. Above all, this study addresses that challenge by developing methods to robustly learn spatial pathological patterns from SEEG, which promises to enhance DRE seizure detection accuracy and guide more precise SOZ localization.

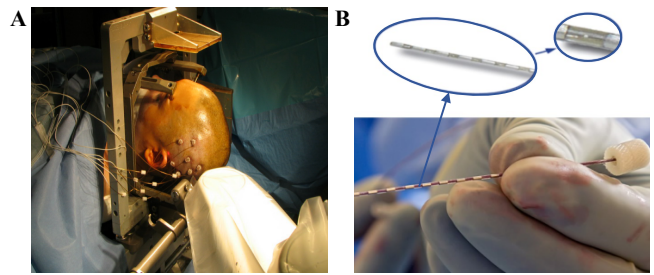

Fig. 1. A. Patients after SEEG surgery. B. Long depth SEEG electrode.

Graph neural networks (GNNs) offer a flexible framework for modeling the complex non-Euclidean spatial network directly from SEEG[11, 12]. By treating each electrode contact as a node and representing directed functional connectivity as weighted edges, GNNs embed both local signal features, such as spikes and high-frequency oscillations, and the spatial interactions among contacts into a graph structure. In studies using non-invasive scalp EEG, GNNs have shown considerable promise in automatically extracting higher-order connectivity features and in classifying brain states or disorders[13, 14], such as sleep stage classification and emotion recognition. However, two major challenges have limited GNN-based seizure detection from SEEG of DRE patients: (1) The low signal-to-noise ratio of connectivity estimates caused by volume conduction, physiological artifacts and estimation errors, making it difficult to learn seizure-related edges when propagated through GNNs. (2) The absence of expert labels for both interictal suppression and ictal propagation patterns complicates the supervised learning of these essential spatial features.

First, to prevent unrelated edges from being amplified when aggregating over a GNN, much related work has been done in recent years. Liu et al. [15] proposed the Graph Contrastive Denoising (GCD) module, which uses contrastive pretraining to filter out irrelevant edges and reinforce class-relevant connections. Zhang et al. [16] and Zhu et al. [17] introduced topology-aware enhancement strategies that remove low-importance nodes and their edges with higher probability, thereby preserving the core structure. Suresh et al. [18] jointly train the enhancer parameters and the GNN encoder end-to-end using a single contrastive loss, which significantly improves the adaptability of the enhancement strategy. These studies have provided insights from multiple perspectives, yet a comprehensive joint approach is still needed for SEEG seizure analysis. Based on this motivation, we address the SNR problem in edge connections by learning a clearer graph structure that simultaneously strengthens epileptogenic topologies and denoises irrelevant edges.

Secondly, contrastive learning (CL) is a powerful technique for extracting discriminative features from limited labels[19, 20]. In particular, self-supervised and unsupervised CL methods exploit unlabeled recordings by attracting similar examples and repelling dissimilar ones in the embedding space without labels[21]. This method provides key inspiration for characterizing unlabeled SEEG spatial patterns. Zhao et al. [22] showed that contrastive objectives implemented via time and frequency domain augmentations can produce more invariant and transferable representations, thereby reducing reliance on labeled datasets. Li et al. [23] minimized the distance between pairs of homologous transformations while simultaneously maximizing the distance between pairs of heterologous transformations. Guo et al. [24] learned intrinsic epileptic EEG patterns across subjects by contrastive learning. Motivated by these advances, we integrate a dual-view contrastive framework to exploit unlabeled SEEG patterns and learn robust spatial pathological features.

To overcome the limitations in prior research and represent SEEG spatial pathological patterns, we propose Seizure-NGCLNet, a novel framework based on node-graph dual contrastive learning, as shown in Fig. 2. We aim to learn unlabeled spatial pathological patterns through the supervised task of seizure detection, thereby improving the accuracy of seizure detection. To our best knowledge, this is the first work to integrate dual contrastive learning in SEEG epilepsy detection. Our contributions are as follows:

- Adaptive Graph Augmentation: We employ graph centrality-driven perturbations to selectively modify high-importance edges, generate noise-robust graph variants that preserve critical structure.
- Dual-Contrastive Learning: We integrate global graph-level and local node-graph contrast objectives to denoise and enrich comprehensive SEEG representations. A top-k localized graph attention network is used to fine-tune to perform the final detection.
- State-of-the-Art Seizure Detection: Evaluated on SEEG of 33 DRE patients, our method outperforms baselines and existing contrastive models in accuracy and sensitivity.
- Clinical Interpretability: The learned embeddings reveal two connectivity patterns, interictal suppression and ictal propagation, aligning with clinical mechanisms.

TABLE I. PATIENT DEMOGRAPHIC

|  | *Pre-selection* | *Post-selection* |
|---|---|---|
| Number of patients | 57 | 33 |
| Gender, male/female | 27/30 | 17/16 |
| Age onset, years | 16.51±12.45 | 14.79±11.09 |
| Age surgery, years | 34.07±10.42 | 32.09±10.62 |
| Surgical outcome, Engel 1/2/3/4 | 36/7/12/2 | 19/5/9/0[a] |
| Sampling rate | 256/500/512/1024 | 500/512/1024 |
| Electrode ECOG/SEEG | 20/37 | 0/33 |

[a.] Select patients with Engel 1 ABC, Engel 2A and 3 A

## II. MATERIALS AND METHODS

### A. Dataset and Data preprocessing

*1) Dataset:* The data used in this study is obtained from the Hospital of the University of Pennsylvania (HUP) iEEG epilepsy dataset (https://openneuro.org/datasets/ds004100) [6, 25, 26], which includes data from 57 drug-resistant epilepsy patients who underwent intracranial EEG with subdural grid, strip electrodes or depth electrodes. Each patient also underwent subsequent received treatment with surgical resection or laser ablation. Electrophysiologic data for both interictal and ictal periods are available, along with electrode localizations in the ICBM152 MNI space. Each patient's implantation scheme varies, with between roughly 100 to 200 electrode contacts placed in brain regions suspected to be involved in seizures (covering lobes like temporal, frontal, etc.). SEEG signals are recorded and digitized at 500Hz or 1024Hz sampling rates using a referential montage and preprocessed to eliminate line noise. Electrode configurations and signal labels are determined by a multidisciplinary team of neurologists and neurosurgeons. The data were collected retrospectively under appropriate Institutional Review Board approvals, and de-identified for research use.

We select and analyze 33 patients who received SEEG investigations followed by epilepsy resection or ablation surgery, as shown in TABLE I. The patient inclusion criteria are as follows: (a) absence of recorded seizures or sampling rate below 500 Hz; (b) seizure duration less than 20s; (c) a minimum of six months of post-operative follow-up; (d) at least one recorded seizure exhibiting rhythmic activity.

*2) Data processing:* The SEEG dataset consists of interictal (non-seizure) and ictal (seizure) signals, and is provided in BIDS format, which facilitates standardized data organization and processing. Given the seizure onset time, the data that 10 s before and after each seizure onset time have been selected for further analysis. To improve the discriminative information, we remove the bad channels, and segment the data into 2-second for further analysis. All data processing and analysis are performed using MATLAB R2021a, and classification and prediction models are built using Python3.11.

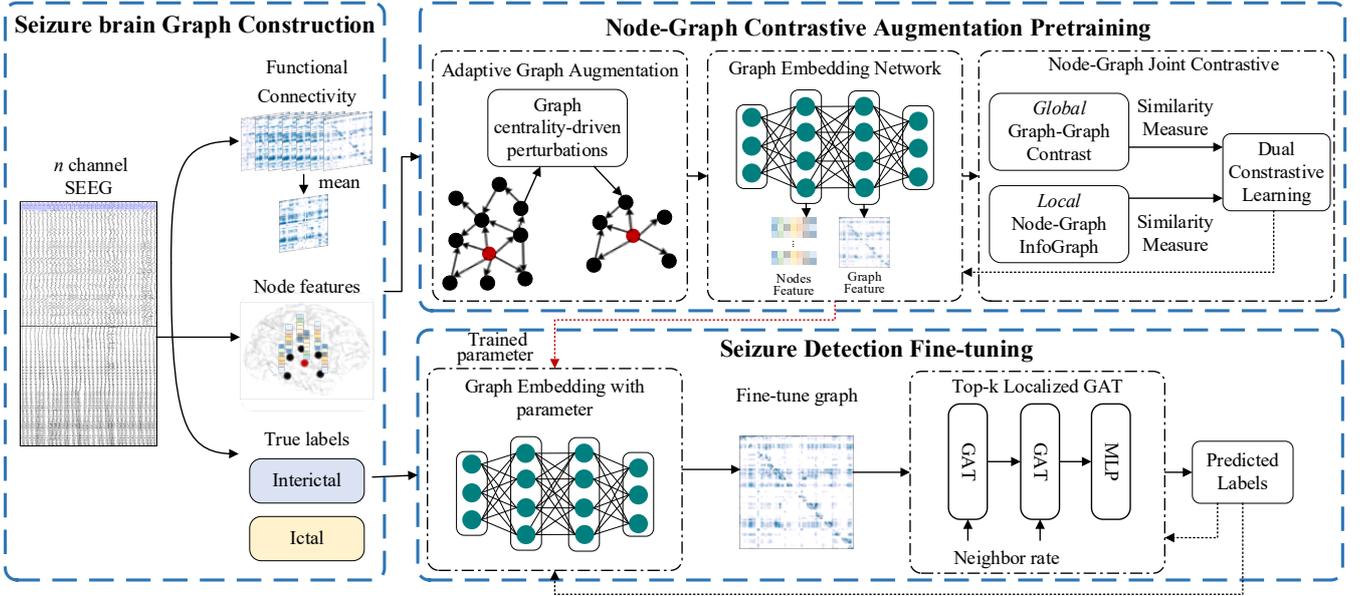

Fig. 2. The overall architecture of our proposed Seizure-NGCLNet. First, in seizure brain graph construction module, we describe SEEG as a directed graph $\mathcal{G} = (V, E, W)$. Second, in node-graph contrastive augmentation pretraining module, adaptive graph augmentation strategy guided by centrality metrics is developed to generate seizure-related brain networks. A dual-contrastive learning approach is integrated, combining global graph-level contrast with local node-graph contrast, to encode both spatial structural and semantic epileptogenic features. Third, in seizure detection fine-tuning module, the pretrained embeddings are fine-tuned via a top-k localized graph attention network to perform the final classification.

*B. Problem Definition and overview*

We regard epileptic seizure detection as a graph-level classification problem. Given a brain graph $\mathcal{G}$ constructed from an SEEG segment, the task is to predict a binary label $\hat{Y}$ indicating whether a seizure is present in that segment. During training, we have a set of graphs for each patient, with $Y = 1$ for ictal graphs and $Y = 0$ for interictal graphs. The goal is to learn a model $F$ that produces a discriminative representation (embedding) for each graph, which can be fed to a classifier to output $\hat{Y}$:

$$\hat{Y} = F(\mathcal{G}) \quad (1)$$

We adopt a self-supervised pre-training and fine-tuning paradigm, the overall architecture contains three main components: (1) a seizure brain graph construction module, (2) a node-graph contrastive augmentation pretraining module, and (3) a seizure detection fine-tuning module based on GAT.

*C. Seizure brain Graph Construction*

We construct seizure-related brain graphs from SEEG, representing the brain as a directed functional network. Specifically, each SEEG electrode contact is represented as a node, with directed edges between nodes indicating the causal relationships inferred from functional connectivity analysis. The graph for each SEEG segment $t$ can thus be formally described as a directed graph $\mathcal{G} = (V, E, W)$, where the set $V$ corresponds to SEEG contacts, the set $E$ contains directed connections among these nodes, and $W$ is a weighted connectivity matrix representing the strength and direction of the connections.

*1) Directed Functional Connectivity based on Multi-band Directed Transfer Function:* To accurately represent directed interactions between brain regions, we utilize the Directed Transfer Function (DTF), applied separately across multiple standard SEEG frequency bands, such as delta, theta, alpha, beta, gamma, ripple, and fast ripple. The DTF measures effective connectivity by quantifying causal influence among electrode contacts.

The SEEG data segments are divided into 2s analysis windows $T$ with 50% overlap. Within each time window, we first fit a multivariate autoregressive (MVAR) model to estimate the relationship between channels. The MVAR model is defined mathematically as follows:

$$\sum_{k=0}^{P} \Lambda(k) S(t-k) = E(t) \quad (2)$$

where $p = 10$ is chosen by the Akaike Information Criterion (AIC).

MVAR parameters are transformed into frequency-domain coefficients $H(f)$, encoding directional influences. The connectivity strength from node $j$ to node $i$ at frequency $f$ is defined as:

$$\theta_{ij} = |H_{ij}(f)|^2 \quad (3)$$

Integrating across frequencies $f_1$ to $f_2$:

$$\phi_{ij}(f_1, f_2) = \sum_{f=f_1}^{f_2} \psi_{ij}(f) \quad (4)$$

Edges below the top 25% threshold are set to zero, removing weak connections. The final connectivity graph is the average across all frequency bands.

*2) Node Feature Representation based on Epileptogenicity Index:* In addition to connectivity structure, the seizure brain graph nodes carry important clinical information related to epileptogenicity. To encode the epileptic nature of each SEEG contact, we calculate a comprehensive set of epilepsy-related biomarkers and numerical features. Specifically, the node features include occurrence rates of typical epileptiform events and statistical measures that reflect epileptic intensity.

Node features represent epileptogenic strength at each electrode and include: Spike occurrence rate, detected using a validated SEEG-Net model, which has verified by the author.

High-frequency oscillation (HFO) occurrence rate, detected using Hilbert envelope methods. And numerical epileptogenic biomarkers, such as Sample entropy, Petrosian fractal dimension (PFD), Katz fractal dimension (KFD).

Thus, these features form a node feature matrix characterizing epileptic activity comprehensively. The resulting brain graphs effectively combine connectivity structure and epileptogenic node properties, facilitating downstream dual-contrastive learning for robust seizure detection.

### D. Node-Graph Contrastive Augmentation Pretraining

*1) Adaptive Graph Augmentation:* To robustly capture meaningful spatial epileptic patterns and improve model generalization, we propose an adaptive graph augmentation approach. The augmentation procedure intelligently perturbs graph structures according to node importance metrics. Nodes with lower centrality, which typically contribute less to seizure identification, are more likely to experience perturbations, ensuring essential seizure-related information remains intact.

Given a batch of adjacency matrices, firstly, we calculate node importance metrics of degree denoted as $c_i$ for graph $\mathcal{G}_i$. We then perform min-max normalization on the node importance scores within each graph to derive normalized importance $c_i^{norm}$.

Secondly, we probabilistically mask nodes based on their normalized importance. Specifically, lower-importance nodes have a higher probability of being masked. From this probability distribution, we sample nodes for masking. After node selection, all edges connected to masked nodes are removed by setting corresponding rows and columns.

Thirdly, to encourage the model to learn stable representations across varied structural conditions, we apply random perturbations to edges connecting node pairs with lower average importance. The importance of an edge between nodes in graph $\mathcal{G}_i$ is calculated by averaging their node importances. We generate perturbation probabilities for each edge based on inverted edge importance, where the summation is taken over all upper-triangular edge pairs. Subsequently, we randomly select edges to perturb.

Through these adaptive perturbation procedures, our method generates meaningful augmented graph views that preserve critical seizure-related structural information while promoting robustness and generalizability of the learned representations.

*2) Graph Embedding Network:* We propose a graph embedding network to encode node and graph. It captures local epileptogenic characteristics at each node, and summarizes global connectivity features and implicitly filters noise, producing robust global graph embeddings.

Given an augmented adjacency matrix $A$ and associated node features $W$, the network produces two complementary embeddings.

The node embedding process is defined as follows:

$$H_{node} = f_{\theta_{node}}(A, W), \quad H_{graph} = f_{\theta_{graph}}(H_{node}) \quad (5)$$

where $f_{\theta_{node}}$ represents a multilayer perceptron (MLP) with nonlinear activations (ReLU), and dropout regularization to avoid overfitting. The resulting node embeddings $H_{node}$ is the embedding local seizure-related information at each electrode contact.

Next, the graph-level embedding $H_{graph}$ is computed by aggregating node embeddings. This aggregation employs another MLP.

Additionally, the graph-level embedding decoder reconstructs a refined adjacency matrix, implicitly performing denoising by emphasizing seizure-related connections while suppressing irrelevant connections.

*3) Node–Graph Joint Training Loss:* We introduce a novel joint contrastive learning framework consisting of two losses.

*a) Graph Contrastive Loss for Graph-level Denoising:* Firstly, the graph contrastive loss encourages the embedding model to separate graphs belonging to two classes, while simultaneously bringing embeddings from the same class closer. This loss utilizes two similarity measures: Global Structural Similarity: Captured by the RBF kernel of the Laplacian eigen-spectra, reflecting the global structure of graphs. Local Embedding Similarity: Captured by the cosine similarity of graph embeddings.

We first reconstruct the adjacency matrix from embeddings:

$$L = I - D^{-1/2} \hat{A} D^{-1/2} \quad (6)$$

where $L$ is the normalized Laplacian matrix and $D$ the degree matrix. We then extract the eigenvalues $\lambda$ of $L$:

$$\lambda = eig(L) \quad (7)$$

The global similarity between graph $\mathcal{G}_i$ and graph $\mathcal{G}_j$ is computed using the Radial Basis Function (RBF):

$$S_{\text{global}}(i,j) = exp\left(-\frac{\|\lambda_i - \lambda_j\|^2}{2\sigma^2}\right) \quad (8)$$

where $\sigma$ is a scaling parameter, set empirically based on eigenvalue distribution.

Next, we calculate local embedding similarity via cosine similarity:

$$S_{\text{local}}(i,j) = \frac{H_{\text{graph}}(i) \cdot H_{\text{graph}}(j)}{\|H_{\text{graph}}(i)\| \|H_{\text{graph}}(j)\|} \quad (9)$$

We integrate both similarities into a unified similarity measure:

$$S(i,j) = \gamma \cdot S_{\text{global}}(i,j) + (1-\gamma) \cdot S_{\text{local}}(i,j) \quad (10)$$

where $\gamma$ balances global and local similarities (empirically set at 0.5). We then define the graph-level contrastive loss using an InfoNCE-style formulation:

$$\mathcal{L}_{graph} = -\frac{1}{B} \sum_{i=1}^{B} \log \frac{exp(S(i, i^+)/\tau)}{\sum_{j \neq i} \exp(S(i,j)/\tau)} \quad (11)$$

where $i^+$ is a positive sample from the same class as $i$, and $\tau$ is a temperature hyperparameter controlling the sharpness of the distribution. Minimizing this loss naturally pushes graphs from different seizure states apart while pulling those from the same class closer, implicitly denoising irrelevant edges and retaining seizure-relevant patterns.

*b) InfoGraph Loss for Node–Graph Information Maximization:* The second component, InfoGraph loss, maximizes mutual information between node embeddings and the corresponding graph-level embedding. This enhances the representational power of node embeddings, ensuring they encapsulate meaningful graph-level information.

The InfoGraph loss is computed as an InfoNCE mutual information maximization problem. For each node embedding in graph $\mathcal{G}$, we form positive pairs by pairing it with its own graph embedding $h_\mathcal{G}$, and negative pairs by pairing with graph embeddings from other graphs $h_{\mathcal{G}'}$:

$$\mathcal{L}_{InfoGraph} = -\frac{1}{B \cdot N} \sum_\mathcal{G} \sum_{v \in \mathcal{G}} log \frac{exp(sim(h_v, h_\mathcal{G})/\tau)}{\sum_{\mathcal{G}'} exp(sim(h_v, h_{\mathcal{G}'})/\tau)} \quad (12)$$

where similarity is again defined by cosine similarity:

$$sim(h_v, h_\mathcal{G}) = \frac{h_v, h_\mathcal{G}}{\|h_v\| \|h_\mathcal{G}\|} \quad (13)$$

**Algorithm 1:** The Description of Seizure-NGCLNet.
**Input:** Multichannel SEEG data
**Output:** Binary classification results of the model
1: **Divide** SEEG into 2s windows with 50% overlap. Each window is labeled as seizure (ictal=1) or non-seizure (interictal=0).
2: For each segment t:
3:   a. construct directed functional brain graph $\mathcal{G} = (V, E, W)$;
4:   b. compute node features (spike/HFO rates, SE, PFD, KFD).
5: For each patient, feed graphs $\mathcal{G}$ and binary labels $\hat{Y}_t$ into dataset $D$.
// Node-Graph Contrastive Learning Pretraining
6: **for** each training epoch **do**
7:   **for** each batch of graphs in $D$ **do**
8:     a. generate an augmented graph $A$ using adaptive perturbation of nodes and edges;
9:     b. encode $H_{node}$ and $H_{graph}$;
10:    c. reconstruct the adjacency matrix for denoising;
11:    d. compute graph-level contrastive loss $\mathcal{L}_{graph} = Laplacian + embedding\ similarity$;
12:    e. compute InfoGraph loss $\mathcal{L}_{InfoGraph}$ to maximize mutual information between node and graph embeddings;
13:    f. update parameters by minimizing $\mathcal{L}_{total} = \mathcal{L}_{graph} + \alpha \mathcal{L}_{InfoGraph}$.
14:  **end for**
15: **end for**
// Seizure Detection Fine-tuning based on Top-k Localized GAT
16: **for** each labeled graph $\mathcal{G}$ **do**:
17:    a. extracts the graph embedding using the pretrained encoder;
18:    b. feed the embedding into the GAT classifier and output seizure probability $\hat{Y}_t$;
19: **end for**
20: Return seizure detection predictions and attention-based interpretability scores from the GAT.

TABLE II. DETAILED RANGE OF PARAMETERS AND FINAL VALUE

| | Parameters | Range | Final |
|---|---|---|---|
| Pretraining | Hidden layers | [16,32,64,128,256] | 256 |
| | Learning Rate | [1e−2,1e−3,1e−4] | 1e−3 |
| | Weight Decay | [0,1e−3,1e−4] | 0 |
| | Node mask ratio | [0,0.2,0.5,0.7] | 0.2 |
| | Edge perturb ratio | [0,0.2,0.5,0.7] | 0.2 |
| | Margin of loss | [0.3,0.5,0.7] | 0.3 |
| | Num of epoch | [40,50,100] | 50 |
| | Loss $\alpha$ ratio | [0.3,0.5,0.7,1] | 1 |
| GAT Fine-tuning | Num of Layers | [1,2,3,4] | 2 |
| | Embedding Size | [8,16,32] | 16 |
| | Neighbor rate | [0,0.3,0.5,0.7,1] | 0.5 |
| Fine-tuning | Learning Rate | [1e−2,1e−3,1e−4] | 1e−3 |
| | Weight Decay | [1e−3,1e−4,1e−5] | 1e−4 |
| | Batch Size | [16,32,64,128,256] | 128 |
| | Num of epoch | [40,50,100] | 50 |

*c) Node–Graph Joint Training Loss:* Our final contrastive training loss combines these two losses into a unified joint training objective:

$$\mathcal{L}_{total} = \mathcal{L}_{graph} + \alpha \mathcal{L}_{InfoGraph} \quad (14)$$

Here, the parameter $\alpha$ balances the node-graph mutual information maximization and edge-level denoising tasks. Jointly optimizing this objective allows the embeddings to capture robust, denoised, and class-discriminative representations suitable for accurate seizure detection.

Through this comprehensive dual-loss strategy, our Seizure-NGCLNet learns highly discriminative node- and graph-level embeddings, effectively distinguishing ictal from interictal states and enhancing seizure-detection accuracy in downstream tasks.

### E. Seizure Detection Fine-tuning based on Top-k Localized GAT

The final seizure detection classifier directly uses the graph embeddings learned in the pretraining stage by employing a multi-head Graph Attention Network (GAT) layer augmented with top-$k$ localized attention.

To suppress unrelated connections and concentrate on the most informative local structure, we retain only the top $k = neighbor\_rate \times N$ scores per node, zeroing out all others before the SoftMax normalization. This hard pruning enforces sparsity in the attention graph and reduces computational overhead.

All head outputs are concatenated to form the updated node features, which are then fed into subsequent GAT layers or a readout operation. After the final layer, we perform mean pooling over nodes to yield a graph-level representation, followed by a linear output layer that predicts the seizure probability. Training minimizes binary cross-entropy loss on labeled SEEG segments.

This top-k localized attention clustering not only enhances robustness against noisy, non-epileptogenic edges but also yields highly interpretable attention maps: the retained top-k connections directly indicate the most salient node interactions driving the seizure classification decision. Algorithm 1 details the complete fine-tuning procedure.

## III. RESULTS

### A. Experimental settings

The SEEG data segments are partitioned into 2-second analysis windows with 50% overlap. After sliding-window processing, each patient yielded on average approximately 970.3 ictal segments and 596.7 interictal segments, each comprising 1,000 or 2,048 data points. For each subject, we adopt a 10-fold cross-validation strategy: the dataset is divided into multiple non-overlapping folds, and in each iteration, one fold is designated as the test set while the remaining folds are used for model training. This procedure ensures robust evaluation of model generalization within each subject. Hyperparameters are optimized through extensive comparative experiments, with the search ranges and final values reported in TABLE II. Model training is performed using the Adam optimizer. All experiments are implemented in PyTorch and executed on an NVIDIA 3090 GPU.

We use accuracy, sensitivity, and specificity. These metrics provide a comprehensive evaluation, capturing both its ability to correctly identify seizures (sensitivity) and to avoid false positives (specificity). The detailed evaluation matrix and formula are consistent with our team's recently published work [27].

### B. Classification performance of Seizure-NGCLNet

This section presents the performance of our proposed Seizure-NGCLNet. Fig. 3A illustrates the distributions of accuracy, sensitivity, and specificity obtained via ten-fold cross-validation for all 33 patients. The model achieves a high mean accuracy of 95.93%, indicating robust overall classification; a mean sensitivity of 96.25%, demonstrating excellent performance in correctly identifying seizure events and thereby reducing false-negative predictions significantly. Moreover, specificity averages 94.12%, confirming the model's reliable performance in accurately classifying non-seizure states, thus minimizing false-positive predictions effectively. Fig. 3B shows the ROC curves for patient 142 across the ten folds, with the bold red curve representing the fold-averaged ROC. The model achieves an impressive mean AUC of $0.9882 \pm 0.0049$, underscoring the discriminative ability between ictal and interictal.

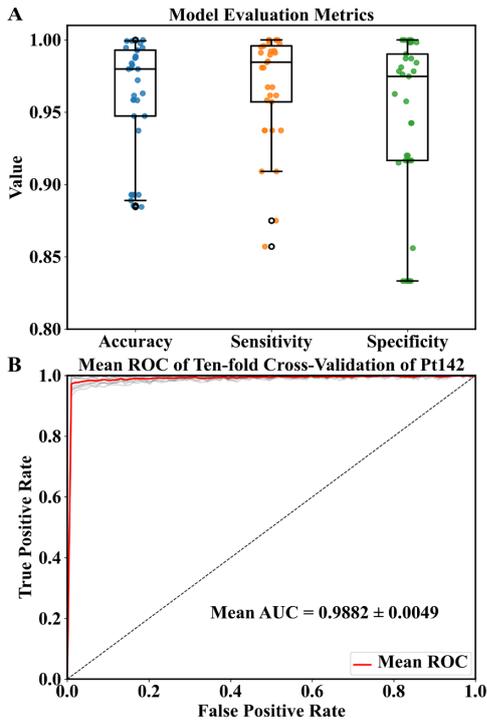

Fig. 3. Classification performance of the Seizure-NGCLNet from seizure detection. A: Distribution of evaluation indicators in ten-fold cross-validation for all 33 patients. B: ROC curve of the training process for patient 142.

The results can be attributed to our novel adaptive graph augmentation strategy, dual contrastive learning approach, and the effective integration of GAT fine-tuning based on Top-k localized attention. These components enable the model to capture critical epileptogenic spatial patterns in different states and suppress irrelevant connections, significantly enhancing seizure detection accuracy and reliability.

*C. Baselines and Ablation study*

To evaluate the effectiveness of each component in the Seizure-NGCLNet, we conducted extensive comparisons with baseline models and performed ablation experiments. As shown in Fig. 4, eight baselines and ablated variants are tested:
- A: 1D-CNN.
- B: Bi-LSTM.
- C: GNN without Contrastive Pretraining.
- D: GAT without Contrastive Pretraining.
- E: Seizure-NGCLNet without Adaptive Augmentation.
- F: Seizure-NGCLNet without Graph Contrastive Loss.
- G: Seizure-NGCLNet without InfoGraph Loss.
- H: Seizure-NGCLNet without Top-k Attention.

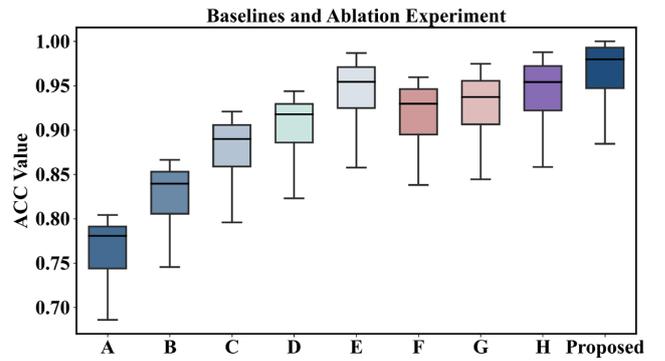

Fig. 4. The boxplot illustrates the ACC distributions for different model variants.

Our Seizure-NGCLNet achieved the highest performance across all evaluation metrics, significantly outperforming the baseline and ablation approaches. In contrast, variants A and B demonstrate the discriminative ability of spatial graph modeling over temporal models. C and D show that removing contrastive pretraining leads to significant drops in performance, with accuracy falling to 87.57%, confirming the benefit of self-supervised graph representation learning.

Further analysis of ablation variants highlights the necessity of each innovation: Removing adaptive graph augmentation (variant E) reduces mean accuracy to 94.1%, as the model becomes less robust to noise and less capable of learning stable representations across diverse patients and recording conditions. The absence of the graph contrastive loss (variant F) results in 4.3% decline in the model's ability to distinguish between ictal and interictal connectivity patterns. This results underscores the importance of contrastive learning for enhancing the separability of pathological and non-pathological brain states in SEEG graph representations. Without InfoGraph node–graph mutual information maximization (variant G) results in mean ACC 92.5%, weaker integration of local pathological node features into the global graph representation. Variant H shows the ability to highlight seizure relevant interactions, decreasing both accuracy and interpretability. In summary, the ablation results unequivocally show that each module in Seizure-NGCLNet is indispensable for maximizing detection performance and interpretability.

*D. Comparison with State-of-the-Art Algorithms*

To evaluate the effectiveness of Seizure-NGCLNet, we conducted comparative experiments on 33 DRE patients from the OpenNeuro SEEG dataset. All models were trained and tested under identical settings to ensure fair evaluation. Table X summarizes the average performance of Seizure-NGCLNet alongside several advanced approaches, including GNN-RNN-based, GraphCL-based, and multi-view GraphCL-based models such as STGAT-GRU[28], DynSeizureGAT[27], DNGCL[19], GCD[15], and TS-GAC[21].

Seizure-NGCLNet outperformed all competing methods in overall classification accuracy and demonstrated a notable advantage in sensitivity. The GNN-RNN-based models achieved lower accuracy about 90.9%-92.17% and sensitivity about 92.3%-94.51%, whereas GraphCL-based methods delivered superior results, reflecting their capacity to capture relevant spatial features in different brain states. Additionally, the multi-view GraphCL-based models exhibited further improvements across multiple metrics, suggesting that temporal dynamics and multi-scale spatial patterns were more effectively represented.

TABLE III. COMPARISON WITH OTHER ALGORITHMS

| Algorithms | Graph Feature | Classifier | ACC(%) | SEN(%) | SPE(%) | PPV(%) | NPV(%) |
|---|---|---|---|---|---|---|---|
| STGAT-GRU[28] | Phase locking value | STGAT-GRU | 90.9 | 92.3 | 89.5 | 89.83 | 92.08 |
| DynSeizureGAT[27] | Sparse DTF | STPGAT-TCN | 92.17 | 94.51 | 89.83 | 90.3 | 94.23 |
| DNGCL[19] | GCN with augmentation strategies | MLP | 93.08 | 94.32 | 91.84 | 92.03 | 94.17 |
| GCD[15] | Denoised graph learning | GNN | 94.06 | 94.33 | 93.79 | 93.86 | 94.35 |
| TS-GAC[21] | Multi-Windows Temporal Contrasting | GNN | 94.23 | 94.45 | 94.01 | 94.07 | 94.44 |
| **Proposed** | **Augmentation strategies and Node-Graph Dual Contrastive Learning** | **Top-k GAT** | **95.93** | **96.25** | **94.12** | **94.96** | **96.75** |

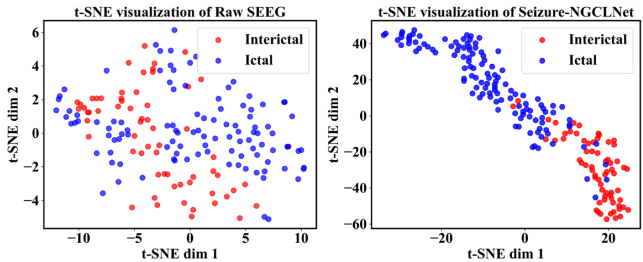

Fig. 5. Two-dimensional t-SNE visualization of feature distribution for both raw SEEG and Seizure-NGCL features.

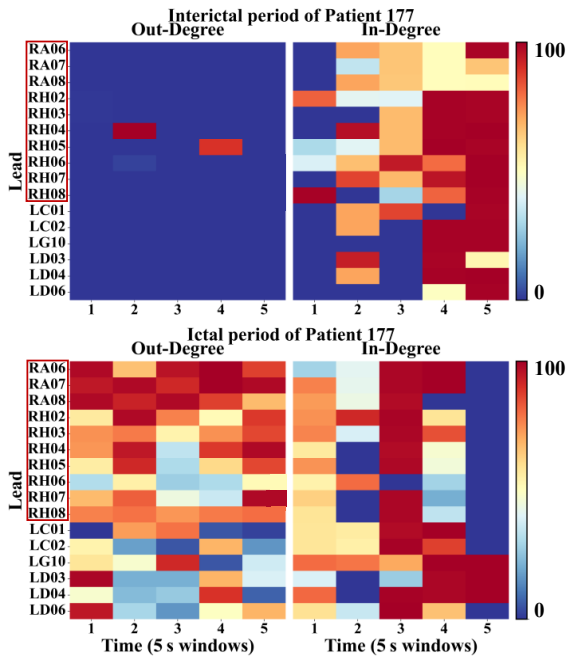

Fig. 6. Nodes' feature heatmap visualizations of patient HUP177.

## IV. DISCUSSION

In this study, we introduced Seizure-NGCLNet, a node-graph dual contrastive framework that decodes SEEG to capture spatial-pathological patterns of DRE and classify interictal/ictal state. We highlight its contributions from two aspects.

First, visualization of the learned graph representation. Fig. 5 presents two t-SNE visualizations of patient HUP148, comparing raw SEEG signals to the graph-level embeddings produced by Seizure-NGCLNet. Each point represents a 2-second SEEG segment, labeled as interictal or ictal. Raw SEEG data shows no clear distinction between states, highlighting their complexity. In contrast, embeddings generated by our model form distinct clusters, clearly separating ictal from interictal segments. This demonstrates that our dual contrastive pretraining combined with GAT fine-tuning effectively captures discriminative spatial features, enabling robust seizure detection.

Second, visualization of the node representation. Fig. 6 presents nodes' feature heatmap visualizations of patient HUP177. Quantitative analysis of attention weights and node saliency maps reveals that the model captures distinctive spatial pathological mechanisms. The interictal state is concentrated on unidirectional flows from SOZ regions, reflecting the suppression patterns, consistent with clinical understanding of epileptic brain dynamics[6, 10]. During the ictal period, the network identifies broad, bidirectional connections reflecting the global diffusion of epileptic activity[10].

Although our results are promising, several limitations remain. First, as previously mentioned, we trained models individually for each patient to address variability in electrode montages. Future research will focus on developing a generalized cross-patient representation learning framework. Second, our pretraining and fine-tuning tasks were both focused on seizure detection. Extending the downstream tasks to epileptogenic contact localization could provide direct clinical utility.

## V. CONCLUSION

We have presented Seizure-NGCLNet, a node–graph dual contrastive learning framework designed to decode complex spatial-pathological patterns—interictal suppression and ictal propagation—from SEEG recordings and to deliver high-precision seizure detection. By (1) employing an adaptive, centrality-guided graph augmentation to denoise connectivity estimates, (2) integrating global graph-level and local node–graph contrastive objectives to learn both structural and semantic epileptogenic features, and (3) fine-tuning with a top-k localized graph attention network, our model achieves state-of-the-art performance on data from 33 DRE patients (95.9% accuracy, 96.3% sensitivity, 94.1% specificity). Seizure-NGCLNet represents interpretable embeddings that cleanly separate ictal from interictal segments, and consistent with clinical mechanism.

ACKNOWLEDGMENTS

This work was supported by National Natural Science Foundation of China (62203063), Beiiing Natural Science Foundation (L242097), Tianjin Municipal Education Commission Scientific Research Project under Grant 2023KJ067. This work was supported by the High-performance Computing Platform of Tianjin Medical University. We thank the NIH, JHH, UMMC, and UMF investigators for providing access to the data used in this study.

REFERENCES

[1] E. Perucca, P. Perucca, H. S. White, and E. C. Wirrell, "Drug resistance in epilepsy," *The Lancet Neurology,* vol. 22, no. 8, pp. 723-734, 2023/08/01/, 2023.

[2] B. Sultana, M.-A. Panzini, A. Veilleux Carpentier, J. Comtois, B. Rioux, G. Gore, P. R. Bauer, C.-S. Kwon, N. Jetté, C. B. Josephson, and M. R. Keezer, "Incidence and prevalence of drug-resistant epilepsy: a systematic review and meta-analysis," *Neurology,* vol. 96, no. 17, pp. 805-817, 2021.

[3] W. H. Organization, "Epilepsy: a public health imperative," 2019.

[4] S. Rheims, M. R. Sperling, and P. Ryvlin, "Drug-resistant epilepsy and mortality—Why and when do neuromodulation and epilepsy surgery reduce overall mortality," *Epilepsia,* vol. 63, no. 12, pp. 3020-3036, 2022/12/01, 2022.

[5] R. Gadot, G. Korst, B. Shofty, J. R. Gavvala, and S. A. Sheth, "Thalamic stereoelectroencephalography in epilepsy surgery: a scoping literature review," *Journal of Neurosurgery,* vol. 137, no. 5, pp. 1210-1225, 2022.

[6] J. M. Bernabei, A. Li, A. Y. Revell, R. J. Smith, K. M. Gunnarsdottir, I. Z. Ong, K. A. Davis, N. Sinha, S. Sarma, and B. Litt, "Quantitative approaches to guide epilepsy surgery from intracranial EEG," *Brain,* vol. 146, no. 6, pp. 2248-2258, Jun 1, 2023.

[7] J. Li, O. Grinenko, J. C. Mosher, J. Gonzalez-Martinez, R. M. Leahy, and P. Chauvel, "Learning to define an electrical biomarker of the epileptogenic zone," *Human Brain Mapping,* vol. 41, no. 2, pp. 429-441, Feb 1, 2020.

[8] F. Bartolomei, S. Lagarde, F. Wendling, A. McGonigal, V. Jirsa, M. Guye, and C. Bénar, "Defining epileptogenic networks: contribution of SEEG and signal analysis," *Epilepsia,* vol. 58, no. 7, pp. 1131-1147, 2017.

[9] D. J. Doss, J. S. Shless, S. K. Bick, G. S. Makhoul, A. S. Negi, C. E. Bibro, R. Rashingkar, A. Gummadavelli, C. Chang, and M. J. Gallagher, "The interictal suppression hypothesis is the dominant differentiator of seizure onset zones in focal epilepsy," *Brain,* vol. 147, no. 9, pp. 3009-3017, 2024.

[10] B. Frauscher, D. Mansilla, C. Abdallah, A. Astner-Rohracher, S. Beniczky, M. Brazdil, V. Gnatkovsky, J. Jacobs, G. Kalamangalam, and P. Perucca, "Learn how to interpret and use intracranial EEG findings," *Epileptic disorders: international epilepsy journal with videotape,* vol. 26, no. 1, pp. 1-59, 2024.

[11] D. Klepl, M. Wu, and F. He, "Graph Neural Network-Based EEG Classification: A Survey," *IEEE Transactions on Neural Systems and Rehabilitation Engineering,* vol. 32, pp. 493-503, 2024.

[12] J. Guo, T. Feng, P. Wei, J. Huang, Y. Yang, Y. Wang, G. Cao, Y. Huang, G. Kang, and G. Zhao, "Adaptive graph learning with SEEG data for improved seizure localization: Considerations of generalization and simplicity," *Biomedical Signal Processing and Control,* vol. 101, pp. 107148, 2025/03/01/, 2025.

[13] J. Wang, X. Wang, X. Ning, Y. Lin, H. Phan, and Z. Jia, "Subject-Adaptation Salient Wave Detection Network for Multimodal Sleep Stage Classification," *IEEE Journal of Biomedical and Health Informatics*, 2024.

[14] Z. Jia, H. Liang, Y. Liu, H. Wang, and T. Jiang, "Distillsleepnet: Heterogeneous multi-level knowledge distillation via teacher assistant for sleep staging," *IEEE Transactions on Big Data*, 2024.

[15] G. Liu, Y. Yan, J. Cai, E. Q. Wu, S. Fang, A. D. Cheok, and A. Song, "GCD: Graph contrastive denoising module for GNNs in EEG classification," *Expert Systems with Applications,* vol. 265, pp. 126013, 2025.

[16] Y. Zhang, H. Zhu, Z. Song, P. Koniusz, and I. King, "Spectral feature augmentation for graph contrastive learning and beyond." pp. 11289-11297.

[17] Y. Zhu, Y. Xu, F. Yu, Q. Liu, S. Wu, and L. Wang, "Graph contrastive learning with adaptive augmentation." pp. 2069-2080.

[18] S. Suresh, P. Li, C. Hao, and J. Neville, "Adversarial graph augmentation to improve graph contrastive learning," *Advances in Neural Information Processing Systems,* vol. 34, pp. 15920-15933, 2021.

[19] P. Jiao, K. Yu, Q. Bao, Y. Jiang, X. Guo, and Z. Zhao, "Graph contrastive learning with node-level accurate difference," *Fundamental Research,* vol. 5, no. 2, pp. 818-829, 2025.

[20] T. Pan, N. Su, J. Shan, Y. Tang, G. Zhong, T. Jiang, and N. Zuo, "GLADA: Global and Local Associative Domain Adaptation for EEG-based Emotion Recognition," *IEEE Transactions on Cognitive and Developmental Systems*, 2024.

[21] Y. Wang, Y. Xu, J. Yang, M. Wu, X. Li, L. Xie, and Z. Chen, "Graph-aware contrasting for multivariate time-series classification." pp. 15725-15734.

[22] C. Zhao, W. Wu, H. Zhang, R. Zhang, X. Zheng, and X. Kong, "Sleep Stage Classification Via Multi-View Based Self-Supervised Contrastive Learning of EEG," *IEEE Journal of Biomedical and Health Informatics*, 2024.

[23] W. Li, H. Li, X. Sun, H. Kang, S. An, G. Wang, and Z. Gao, "Self-supervised contrastive learning for EEG-based cross-subject motor imagery recognition," *Journal of Neural Engineering,* vol. 21, no. 2, pp. 026038, 2024.

[24] L. Guo, T. Yu, S. Zhao, X. Li, X. Liao, and Y. Li, "Clep: Contrastive learning for epileptic seizure prediction using a spatio-temporal-spectral network," *IEEE Transactions on Neural Systems and Rehabilitation Engineering,* vol. 31, pp. 3915-3926, 2023.

[25] A. Li, C. Huynh, Z. Fitzgerald, I. Cajigas, D. Brusko, J. Jagid, A. O. Claudio, A. M. Kanner, J. Hopp, and S. Chen, "Neural fragility as an EEG marker of the seizure onset zone," *Nature neuroscience,* vol. 24, no. 10, pp. 1465-1474, 2021.

[26] J. M. Bernabei, N. Sinha, T. C. Arnold, E. Conrad, I. Ong, A. R. Pattnaik, J. M. Stein, R. T. Shinohara, T. H. Lucas, and D. S. Bassett, "Normative intracranial EEG maps epileptogenic tissues in focal epilepsy," *Brain,* vol. 145, no. 6, pp. 1949-1961, 2022.

[27] Y. Wang, J. Guo, Z. Jia, G. Cao, Y. Yang, G. Kang, and J. Huang, "DynSeizureGAT: Multi-band Dynamic Graph Attention Network for Interpretable Seizure Detection and Analysis of Drug-Resistant Epilepsy Using SEEG," *IEEE Journal of Biomedical and Health Informatics*, 2025.

[28] Y. Wang, Y. Shi, Y. Cheng, Z. He, X. Wei, Z. Chen, and Y. Zhou, "A Spatiotemporal Graph Attention Network Based on Synchronization for Epileptic Seizure Prediction," *IEEE J Biomed Health Inform,* vol. 27, no. 2, pp. 900-911, Feb, 2023.